\documentstyle[12pt]{article}

\newtheorem{dfn}{Definition}[section]
\newtheorem{prp}{Proposition}[section]
\newtheorem{teo}{Theorem}[section]

\newenvironment{axm}{{\bf Axiom}}{}
\newenvironment{prv}{{\bf Proof}}{}
\newenvironment{exm}{{\bf Example}}{}
\newenvironment{clr}{{\bf Corollary}}{}
\newenvironment{note}{{\bf Remark}}{}

\newcommand{\E}{$\cal E$}
\newcommand{\ep}{\varepsilon}

\begin{document}

\begin{center}
 {\large\bf    Synthetic Differential Geometry:\\[7pt]
               A Way to Intuitionistic Models of General Relativity\\
               in Toposes}

\end{center}
\vspace{0.8 truecm}
\begin{center}
        Y.B.~Grinkevich \footnote{email: guts@univer.omsk.su }\\[15pt]
        Department of Mathematics\\
        Omsk State University\\
        644077,Omsk\\
        RUSSIA\\
\end{center}

\begin{center} 31 July 1996 \end{center}
\begin{abstract}
W.Lawvere in \cite{Lawvere_CD} suggested a
approach to differential geometry and to others mathematical disciplines
closed to physics, which  allows to give  definitions
of derivatives, tangent vectors and tangent bundles without
passages to the limits. This approach is based on a idea of consideration
of all settings not in  sets but in some cartesian closed category \E,
particular in some elementary topos.

The synthetic differential geometry (SDG) is the theory
developed by A.Kock \cite{Kock_SDG} in a context of Lawvere's ideas.
In a base  of the theory is an assumption of that a geometric line
is not a filed of real numbers, but  a some nondegenerate  commutative
ring $R$ of a line type in \E.

In this work we shall show that SDG allows to develop a
Riemannian geometry and write the Einstein's equations of a field
on pseudo-Riemannian formal manifold. This give a way for
constructing a intuitionistic models of general relativity
in suitable toposes.
\end{abstract}

\section{Preliminaries}
In this paper will be given some metrical notions in synthetic
differential geometry(SDG).
We shall show that a metrical geometry in SDG is, in general,
similar with a classical one. Most of results will concern
to  so called "global" properties, what means that we will work with
elements aparted from each other.
All notions of SDG are taken from \cite{Kock_SDG}.

As it was shown in \cite{Kock_SDG} the following theory
(a specially Axiom 1)
is not compatible with the axiom of  excluded third so it
have not models in sets but it have so called "well adapted models"
in cartesian closed categories.

Father all settings will be in some cartesian closed category \E.
As it was shown in \cite{Kock_SDG} we can do them using an ordinary
set theoretical language.

As in \cite{Kock_SDG} we shall assume that
a geometric line is a nondegenerate commutative ring $R$ of line type in \E,
i.e satisfies\\
\begin{axm} {\bf 1}
 {\it Let $D =\{ x\in R \ |\ x^2=0 \}$.\\
 For all $g:D\rightarrow R$ are exist the unique $a, b\in R$,
 such, that for all $ d\in D$ is valid $g(d)=a+d\cdot b$.}
\end{axm}
The object $D$ is "generic tangent vector".

To define a metrical notions we have to make some further assumptions
about properties  of $R$.

First of all we shall assume, that on $R$ are given two orders,
agreed with  the structure of the ring:
\begin{enumerate}
  \item the strict order
    $ < $ such, that $\forall x\in R\ \ \ \lnot (x < x) $
  \item the weak order

       $\leq\ $ such, that $\forall x\in R\ \ \ (x\leq x) $
\end{enumerate}
Connected with each other by axioms:
\[ \forall x, y\in R\ \ \ \lnot (x < y)\Rightarrow y\leq x \]
\[ \forall x, y, z\in R\ \ \ x < y \land y\leq z \Rightarrow x < z \]
In a standard manner we shall define intervals.
\[ (x,y)=\{z\in R\ \ |\ \ x < z \land z < y \}     \]
\[ [x,y]=\{z\in R\ \ |\ \ x\leq z \land z\leq y \} \]
We denote by $InvR=\{x\in R\ | \ \exists \ \ y\in R\ \ x\cdot y=1 \} $
-- object of convertible elements in  $R$.\\
We shall assume that the following formula is valid.
\begin{equation}\label{equ_Inv}
  \forall x\in R\ \ \ x\in InvR\ \ \iff\ \ x < 0 \lor x>0
\end{equation}
We shall assume, that:\footnote{Under $\bigwedge_{i=1}^n$ we understand
         $\underbrace{\land \ldots \land}_n $, and under $\bigvee_{i=1}^n$
         we do $\underbrace{\lor \ldots \lor}_n $}
\begin{enumerate}
 \item $R$  is a local ring, i.e
       \begin{equation}\label{equ_local}
          \forall x\in R \ \ \ x\in InvR\ \ \ \lor x-1\in InvR.
       \end{equation}
 \item $R$ is a field of  quotients, i.e
       \begin{equation}\label{equ_fild}
        \forall x_1, \ldots, x_n \in R\ \ \ \lnot (\bigwedge_{i=1}^n x_i=0)
          \Rightarrow \bigvee_{i=1}^n x_i \in InvR.
       \end{equation}
 \item $R$ is a formally real ring i.e
       \begin{equation}\label{equ_freal}
	 \forall x_1, \ldots, x_n \in R\ \ \ \bigvee_{i=1}^n x_i \in
          InvR \Rightarrow \sum_{i=1}^n x_i^2 \in InvR.
       \end{equation}
 \item $R$ is a Pythagorean ring i.e
       \begin{equation}\label{equ_Pyth}
          \forall x_1, \ldots, x_n \in R\ \ \ \sum_{i=1}^n x_i^2 \in InvR
          \Rightarrow \exists \sqrt{\sum\nolimits_{i=1}^n x_i^2}\in InvR.
       \end{equation}
 \item $R$ is a Archimedean ring i.e
       \begin{equation}
	 \forall x\in R\ \ \ x < 0 \lor x < 1 \lor x < 2 \lor \ldots
       \end{equation}
\end{enumerate}
\begin{axm}[Axiom of integration.]\\
   {\it For any $f:[0,1] \rightarrow R$ exists unique $g:[0,1] \rightarrow R$
        such, that $g^\prime \equiv f$ and $g (0)=0$.}\\
\end{axm}
We shall denote $\int\limits_0^1 f (t) dt:=g (1) $.\\
As it is shown in \cite{Kock_SDG},
all these assumptions are realized in well adapted models for $R$.\\
As it is  shown in \cite{McLarty},
from (\ref{equ_local}) and (\ref{equ_fild}) follows that
\begin{equation}\label{equ_main}
   \forall x\in R\ \ \ x<0\ \ \lor\ \ \ (\forall \ep >0\ \ \
   -\ep < x < \ep)\ \ \lor x>0
\end{equation}
We shall denote
\[ R^ +=\{x\in R\ \ |\ \ x>0 \}  \]
\[ R^-=\{x\in R\ \ |\ \ x < 0 \} \]
\[ R^\ep=\{x\in R\ \ |\ \ \forall \ep >0\ \ \ -\ep < x < \ep \}  \]
It is easy to see that (\ref{equ_main}) can be written as follows:
\begin{equation}\label{equ_-+}
 R=R^- \cup R^\ep \cup R^+
\end{equation}

\section{Linear algebra}\label{sec_LA}
As in the basis of our reasonings is the ring $R$ and its properties, for
consideration of a metric we needs in some results from intuitionistic
linear algebra. The initial items of information on this question are
taken from  C.Mulvey "Intuitionistic algebra and
representation of rings"\cite{Mulvey_IA} and  A.Heyting
"Intuitionism"\cite{Heyting_I}, but, as these works
contains only a few results on the theme, some of them
we had to prove.

\vspace{3\baselineskip}
\subsection{Apartness relation on the ring $R$}
The apartness relation in the intuitionistic  mathematics is the
positive form of the not equality relation.
It have been entered and investigated by Heyting
(see for example \cite{Heyting_I}).
In this paragraph we shall give definition of apartness
relation on the ring $R$ and investigate its properties.
It is necessary to notice, that relation given below,
not completely satisfies to Heyting axioms of apartness,
and therefore, we had to check up its properties anew.

\begin{dfn}
{\rm
   We shall speak, that $a, b\in R $ are {\it apart} and  write
   $a\# b$, if $a-b\in InvR$.
}
\end{dfn}
\begin{note}
  From the (\ref{equ_Inv}) follows, that
  \[ a\# b \iff a < b \lor a > b. \]
\end{note}
\renewcommand{\theenumi}{\arabic{enumi}}
\begin{prp}
{\rm The apartness relation on $R$ has following properties}:
 \begin{enumerate}
  \item $ a=b \Rightarrow \lnot (a\# b) $.
  \item $ \lnot (a=b) \iff a\# b $.
  \item $ a\# b \Rightarrow (a\# c) \lor (b\# c)\ \ \ \forall c\in R$.
 \end{enumerate}
\end{prp}
\begin{prv}
 \begin{enumerate}
  \item  Obviously.
  \item  The necessity is follows from (\ref{equ_fild}).\\
	 The sufficiency is obvious.
  \item  $a\# b \Rightarrow a-b \in InvR $ \\
         So as $R$ is local
         ring, we have that for any $x\in R$ and $r\in InvR$
         \[ x\in InvR\ \ \lor\ \ r-x\in InvR.\]
         If we put $r=a-b$ and $x=a-c$, we shall receive
	 \[ a-c \in InvR\ \ \lor \ \ c-b \in InvR. \]
         What means that $a\# c \lor b\# c$. $\Box $
 \end{enumerate}
\end{prv}
 \begin{note}
  Heyting in \cite{Heyting_I} defines an apartness relation,
  as a relation satisfying to conditions:
  \begin{enumerate}
   \item $ a\# b \Rightarrow \lnot (a=b) $.
   \item $ \lnot (a\# b) \Rightarrow a=b$.
   \item $ a\# b \Rightarrow a\# c \lor b\# c\ \ \ \forall c\in R$.
 \end{enumerate}
\end{note}

As we have noticed, the apartness relation on $R$ differs
from what was considered by Heyting, but despite of, for it are also executed
following positive statements.
\begin{prp}\label{prp_OT2}
 {\rm Are executed:}
 \begin{enumerate}
   \item $ a\# b \Rightarrow (a+c)\# (b+c) \ \ \ \forall c\in R$.
   \item $ a\# b, c\# 0 \Rightarrow a c\# b c $.
 \end{enumerate}
\end{prp}
\begin{prv}
 The proof of these statements is based on compatibility of the order $ < $
 with the structure of the ring.
  \begin{enumerate}
   \item $ a\# b \iff (a<b) \lor (a>b) \Rightarrow $
         $ a+c < b+c \lor a+c > b+c \Rightarrow (a+c) \# (b+c) $.
   \item $ a\# b$ and $ c\# 0 \Rightarrow (a<b) \lor (a>b)$ and
         $ c>0 \lor c<0$\\ $c>0$
         $ \Rightarrow (a c < b c \lor a c > b c) $
	  $\Rightarrow a c \# b c $.\\
         $c < 0$ is similar.$\Box $
  \end{enumerate}
\end{prv}
\begin{prp}\label{prp_OT3}
 {\rm Are executed:}
 \begin{enumerate}
   \item $ a\cdot b \# 0 \Rightarrow a \# 0 \land b \# 0$.
   \item $ a+b \# 0 \Rightarrow a \# 0 \lor b \# 0 $.
   \item $ a\cdot b \# c\cdot d \Rightarrow (a \# c) \lor (b \# d) $.
 \end{enumerate}
\end{prp}
\begin{prv}
The proof of this statement is based on the results of
Proposition \ref{prp_OT2} and similar to the appropriate proof in
 (\cite[\S 4.1.3]{Heyting_I}).
\end{prv}
\subsection{Systems of linear equations}
All the theorems below are proven by Heyting \cite{Heyting_I}.
Their proofs are based on positive properties of apartness
relation (Statements \ref{prp_OT2}, \ref{prp_OT3}).

Let $A=(a_{ij})$ be the matrix of a system of linear equations
with  coefficients from $R$.
\begin{equation}\label{equ_SLE}
  \sum_{k=1}^n a_{ik}x_k=b_i \ \ \ (i=1, \ldots, n.)
\end{equation}
Let $d$ be the determinant of $A$.
If $d\# 0$, it is possible to decide the system (\ref{equ_SLE})
using the Cramer's rule
\[  x_k=\frac{d_k}{d}  \]
Decision is unique in a following exact sense:
\begin{teo}[ Heyting \S 4.2.1 ]\label{teo_SLE}
  If $p_1, \ldots, p_n$ are
  such numbers, that for some $j$ takes place $p_j\# d_j / d$,
  then it is possible to find such $i$, that
  \[ \sum_{k=1}^n a_{ik}p_k\# b_i. \]
\end{teo}
\begin{dfn}
{\rm
  A matrix $A$ has a {\it rank} $r$, if at least
  one of it minors  of the order $r$ is apart from a zero,
  while all minors of the order $r+1$ are equal to a zero.
}
\end{dfn}
For a system of similar equations
\begin{equation}\label{equ_OSLE}
  \sum_{k=1}^n a_{ik}x_k=0 \ \ \ (i=1, \ldots, m.)
\end{equation}
we have the theorem.
\begin{teo}[ Heyting \S 4.2.4 ] \label{teo_OSLE}
 If rank of a matrix $A=(a_{ik}) $ is equal $n$, then for any
 $u_1, \ldots, u_n$, such, that $u_k\# 0 $ at least for one $k$,
  will be  though one $i$, such, that
 \[ \sum_{k=1}^n a_{ik}u_k \# 0. \]
\end{teo}
The inverse theorem is valid too.
\begin{teo}[Heyting \S 4.2.4 ] \label{teo_OSLE_1}
 If for any $u_1,\ldots, u_n$, such, that $u_k\# 0 $ and at least for one
 $k$ left-hand part of a system (\ref{equ_OSLE}) is aparted from a zero,
 then rank of $A$ is equal $n$.
\end{teo}

\subsection{$R$-modules with an apartness relation}
In this paragraph we give abstract definition of apartness
relation on $R$-modules, which generalize properties of apartness
relation on $R$  and prove a theorem about dimension of $R$-module's basis.

Let $V$ be $R$-module. We shall name the elements of $V$ as {\it vectors}.
\begin{dfn}\label{dfn_OTV}
{\rm
  Binary relation $\#$ on $V$, satisfying to conditions:
   \begin{enumerate}
     \item $ \bar{a}=\bar{b}\Rightarrow \lnot (\bar{a}\# \bar{b}) $.
     \item $ \lnot (\bar{a}=\bar{b}) \iff \bar{a}\# \bar{b}$.
     \item
    $ \bar{a}\# \bar{b}\Rightarrow \bar{a}\# \bar{c} \lor \bar{c}\# \bar{a}$
   \end{enumerate}
  where $\bar{a} ,\bar{b}, \bar{c} \in V $, will be called  an
  {\it apartness relation} on $V$.
}
\end{dfn}
Further we shall give positive concepts,
equivalent to  classical concepts of linear dependence and linear
independence of vectors. These definitions are given by
analogy to appropriate definitions of Heyting.
\begin{dfn}
{\rm
  Let $V$ be a $R$-module.
  We shall speak, that vectors $\bar{a}_1, \ldots, \bar{a}_m \in  V$
  are {\it strongly linearly dependent}, if exists  $\lambda_i \in R$
  apart from a zero such, that
  \[  \lambda_1 \cdot \bar{a}_1+\ldots+\lambda_m \cdot \bar{a}_m=0  \]
}
\end{dfn}
\begin{dfn}
{\rm
  Let $V$ be a $R$-module with given on it apartness relation.
  We shall speak, that vectors $\bar{a}_1, \ldots, \bar{a}_m \in V$
  are {\it  mutually free }, if from
  that at least one of $\lambda_i$ apart from a zero, follows,
  that
  \begin{equation}
    \lambda_1 \cdot \bar{a}_1+\ldots+\lambda_m \cdot \bar{a}_m \# 0
  \end{equation}
}
\end{dfn}
We shall give the following definition:
\begin{dfn}
{\rm
  Let $V$ be a $R$-module with given on it an apartness relation.
   A system
  of mutually free vectors such, that any vector from $V$ can be expressed as
  a linear combination of vectors of this system  will be called
  a {\it basis} of $V$.
}
\end{dfn}
We shall prove the following theorem by analogy to the classical proof
which is given in the book \cite{Lang_A}, using our definitions and
properties of the ring $R$.

\begin{teo}\label{teo_base}
Let $V$ be a finitely generated $R$-module. Then, any two bases  of
$V$ have identical dimension, that is, contain identical number of vectors.
\end{teo}
\begin{prv}
 Let $\{v_1, \ldots, v_p\}$ be a basis of $V$, $p\geq 1$.

 To prove, that any other basis consists from $p$ of elements, is
 enough to prove that
 if $\{w_1, \ldots, w_r\}$ is a system of mutually free vectors, then
  $r\leq p$. Inverse inequality may be  proven similarly.

 We shall prove by induction.\\
 As $\{v_1, \ldots, v_p\}$ is a basis, the vector $w_1$ can be
 recorded as follows
 \begin{equation}\label{equ_w}
   w_1=c_1 v_1+\ldots+c_p v_p\ \ \ c_1, \ldots, c_p \in R.
  \end{equation}
  As $\{w_1, \ldots, w_r\}$  are mutually free, we have $w_1\# 0$.
 Assumption, that all $c_i=0$, lead to the contradiction.
 Consequently
 \[ \lnot (\bigwedge_{i=1}^p c_i=0)  \]
 Whence, from (\ref{equ_fild}), we shall receive,
  that exists $i$ such, that $c_i\# 0$. We shall consider, for
  definiteness,  that it is $c_1\# 0$.
  Then $v_1$ lies in the space generated by
  $\{w_1, v_2, \ldots, v_p\}$, which coincides with all $V$.

 We shall show, that the vectors $\{w_1, v_2, \ldots, v_p\}$ are mutually
 free.
 Actually, if we shall consider the linear combination
 $\lambda_1 w_1+\sum_{i=2}^p \lambda_i v_i $, such, that at least one
 $\lambda_i \# 0$,
 using (\ref{equ_w}), we shall receive
 \begin{equation}\label{equ_v}
   \lambda_1 c_1 v_1+\sum_{i=2}^p (\lambda_i+c_i \lambda_1) v_i.
 \end{equation}
 In a force  of (\ref{equ_main}), for $\lambda_1 $ we
 have two cases:
 \begin{enumerate}
   \item $\lambda_1 \# 0 $. In this case $\lambda_1 \cdot c_{1} \# 0$ as
          $c_1$  also apart from a zero.
   \item $\lambda_1 \in R^\ep $. In this case $\lambda_1 \cdot c_{i} $ also
         belongs to $R^\ep $, and, hence,
         $\lambda_i+c_i \lambda_1 \# 0$.
 \end{enumerate}
 In any case, from  the mutual freedom of vectors $\{v_1, \ldots, v_p\}$,
 we receive, that the linear combination (\ref{equ_v}) is aparted
 from a zero, and, hence, the vectors $\{w_1, v_2, \ldots, v_p\}$  are
 mutually free.

 We shall assume on a induction, that after suitable renumbering of $v_i$ we
 have found $w_1, \ldots, w_k\ \ (k < p) $ such, that
 $\{w_1, \ldots, w_k, v_{k+1}, \ldots, v_p \}$ is a basis of $V$.
 We shall present  $w_{k+1}$ as follows
 \[  w_{k+1}=c_1 w_1+\ldots+c_k w_k+c_{k+1} v_{k+1}+\ldots c_p v_p, \]
 where at least one $c_i$ apart from a zero.
 We shall assume, that it $c_{k+1}\# 0$.
 Using a similar reasons, we shall change $v_{k+1}$ on $w_{k+1}$
 and again receive a basis. We shall repeat this procedure
 till  $k$ became equal to  $r$.
 Whence we have, that $r\leq p$.\ Hence the theorem is proved.$\Box $
\end{prv}

\subsection{Algebraic properties of the $R^n$}
In this paragraph we shall give definition of an apartness relation on $R^n$
and a notion of basis of $R^n$ .

First of all let us note, that $R^n$ is an $R$-module
in a natural way.
\begin{dfn}
{\rm
  A vectors  $\bar{a}, \bar{b} \in R^n $ are {\it apart},
  $\bar{a}\# \bar{b}$, if $a_i$ and $b_i$ are apart in $R$, at least for one
  $i$.
}
\end{dfn}
\begin{prp}\label{prp_RnOT}
{\rm
  The relation introduced above is really the apartness
  relation on $R^n$ in the sense of Definition~\ref{dfn_OTV}.
}
\end{prp}
\begin{prv}
Let us check up conditions 1 -- 3 of specified definition.
\begin{enumerate}
  \item Obviously.
  \item  The necessity is follows from (\ref{equ_fild}).\\
	 The sufficiency is obvious.
  \item Let $\bar{a}\# \bar{b}$, hence $a_i \# b_i$, at least for one $i$.
        Let $\bar{c}\in R^n $, then from the  properties of apartness in
         $R$ we have
	 \[ a_i \# c_i \ \lor \ c_i \# b_i. \]
         Whence we receive, that
         $\bar{a} \# \bar{c} \lor \bar{c} \# \bar{a}$.$\Box $
\end{enumerate}
\end{prv}

Further we shall give the theorem from Heyting, which  describe
properties of mutually free vectors in $R^n$.
\begin{teo}[ Heyting \S 4.3.1 ] \label{teo_FV1}
 For vectors
 \[
  \bar{a}_i=(a_{i1}, \ldots, a_{in}) \ \ \ (i=1,2, \ldots, p)
 \]
  were mutually free, is necessary and sufficiently, that the matrix
  made from their coefficients had rank $p$.
\end{teo}

Let us consider  a system  of $n$ mutually free vectors
$\bar{e}_1, \ldots, \bar{e}_n$ in $R^n$. Then the matrix made from
coefficients of these vectors, under the Theorem \ref{teo_FV1}, has rank
$n$ and  its determinant apart from a zero. From the Theorem \ref{teo_SLE}
follows, that any vector from $R^n$ can be expressed as a linear combination
of vectors $\bar{e}_1, \ldots, \bar{e}_n$.
>From the Theorem \ref{teo_base} follows, that any basis of $R^n$
consists from $n$ mutually free vectors.

Let
$\{\bar{e}_1, \ldots, \bar{e}_n\}$ and $\{\bar{f}_1, \ldots ,\bar{f}_n \}$
are two bases in $R^n$.
Lay out vectors of the second basis through the first.
\begin{eqnarray*}
  \bar{f}_1    &= & c^1_1 \bar{e}_1+\ldots +c^n_1 \bar{e}_n      \\
  \ldots       &   &   \ldots \ \ \ \ \ \ \ \ \ \ldots              \\
  f_j          &= & c^1_j \bar{e}_1+\ldots +c^n_j \bar{e}_n      \\
  \ldots       &   &   \ldots \ \ \ \ \ \ \ \ \  \ldots             \\
  \bar{f}_1    &= & c^1_n \bar{e}_1+\ldots +c^n_n \bar{e}_n
\end{eqnarray*}
The matrix $C=(c^i_j)$, made from coefficients of decomposition, is
{\it the matrix of transition} from basis
$\{\bar{e} _1, \ldots, \bar{e}_n\}$ to basis
$\{\bar{f}_1, \ldots, \bar{f}_n \}$.
The matrix $C$ is convertible and its determinant  apart  from a zero.

Let $\bar{x}\in R^n $ in a basis
$\{\bar{e}_1, \ldots,\bar{e}_n\}$ has the form
\[ \bar{x}=x^1 \bar{e}_1+\ldots+x^n \bar{e}_n \]
And in basis $\{\bar{f}_1, \ldots, \bar{f}_n \}$:
\[  \bar{x}=y^1 \bar{f}_1+\ldots+y^n \bar{f}_n  \]
Then the coordinates of $x$, in the new and in the old bases,
are connected among themselves as follows:
\[
 x^i=\sum_{j=1}^n c^i_j y^j \ \ \ (i=1, \ldots, n).
\]

\subsection{The space of linear forms}
In this paragraph we will give a definition of apartness relation and
of basis of the space of linear forms.

\begin{dfn}
{\rm
  A {\it linear form} on $R^n$ is
  a map $f:R^n\rightarrow R$ such, that
 \[ \forall  r\in R\ \ \forall \bar{x}\in R^n:\ f(r\cdot \bar{x})
                                                       =r\cdot f(\bar{x}) \]
 \[ \forall \bar{x},\bar{y}\in R^n\ \ f(\bar{y}+\bar{x})=
	f(\bar{x})+f(\bar{y}) \]
}
\end{dfn}
\begin{dfn}
{\rm
  The {\it space of all linear forms} $R^{n*}$ is
  the subobject of $R^{R^n}$  with a structure of $R$-module on it given by
  formulas
  \[ (f+g)(\bar{x})=f(\bar{x})+g(\bar{x})\]
  \[ (r\cdot f)(\bar{x})=r\cdot f(\bar{x})  \]
}
\end{dfn}
Thus  $R^{n*}$ has a structure of $R$-module.

Let $\{\bar{e}_1, \ldots, \bar{e}_n\}$ be a basis of $R^n$.
We shall define the linear forms $f^i$ as follows:
\[ f^i(\bar{e}_j)=\delta^i_j. \]
Obviously, an any linear form $h$ can be expressed as a linear combination of
$f^i$'s:
\[ h (\bar{x})=\sum_{i=1}^n h_i f^i (\bar{x}), \] where
$h_i=h (\bar{e}_i) $.

\begin{dfn}
{\rm We shall speak, that $f\in R^{n*}$ {\it is aparted from a
 zero(linear form) } and write $f\# 0$, if exists $i$ such, that
 $f(\bar{e}_i)\# 0$, where $\{\bar{e}_1, \ldots, \bar{e}_n \}$ is basis of
  $R^n$.\\ $f\# g \iff f-g\# 0$.
}
\end{dfn}
\begin{note}
  This definition does not depend on choice of basis.
  Really, let $\bar{e}_i $ be such basis vector
  that $f(\bar{e}_i) \# 0 $ and
  \[ \bar{a}_i=c_1 \bar{h}_1+\ldots+c_n \bar{h}_n \]
  in a new basis $\{\bar{h}_1, \ldots, \bar{h}_n \}$.
  Then we have
  \[ f(\bar{e}_i)=c_1 f(\bar{h}_1)+\ldots +c_n f(\bar{h}_n)\]
  Whence from the Proposition \ref{prp_OT3} follows that $f(\bar{h}_j) \# 0$.
\end{note}
\renewcommand{\theenumi}{\arabic{enumi}}
\begin{prp}
{\rm
    The relation introduced above is the apartness relation in $R^{n*}$
    in the sense of Definition~\ref{dfn_OTV}.
}
\end{prp}
\begin{prv}
  Proof is similar to the proof of Statement \ref{prp_RnOT}.$\Box $\\
\end{prv}
We have that $R^{n*}$ is a $R$-module with apartness relation,
hence, a notions of a mutual freedom and of a basis are defined on $R^{n*}$
and the theorem \ref{teo_base} is valid.

Also valid the following theorem
\begin{teo}\label{teo_Rank}
{\rm
   The linear forms $g^1, \ldots, g^p \in R^{n*}$ are  mutual free  iff
   rank of a matrix
   \[ \left(
    \begin{array}{ccc}
      g^1(\bar{e}_1) & \ldots     &  g^1(\bar{e}_n)   \\
      \ldots          & \ldots     &  \ldots     \\
      g^p(\bar{e}_1)  & \ldots     &  g^p(\bar{e}_n)
    \end{array}
   \right) \]
  is equal to $p$.
}
\end{teo}
\begin{prv}
  The direct consequence of the Theorems \ref{teo_OSLE}, \ref{teo_OSLE_1}.
  $\Box $\\
\end{prv}
\begin{clr}
System of linear forms $f^i$ is basis of $R^{n*}$, which is {\it dual}
to the basis $\{\bar{e}_1, \ldots, \bar{e}_n \}$.
\end{clr}

Let $\{\bar{e}_1, \ldots, \bar{e}_n \}$ be a basis of $R^n$ and
$\{f^1, \ldots, f^n\}$ its  dual basis of $R^{n*}$.
Let  $y\in R^{n*}$ and
\[ y=y_1 f^1+\ldots+y_n f^n \]
Let $\{\bar{h}_1, \ldots, \bar{h}_n \}$ be another basis of $R^n$ and
$\{j^1, \ldots, j^n\}$  its dual basis of $R^{n*}$. Let
\[ y=z_1 j^1+\ldots+z_n j^n \]
Then the coordinates of $y$ in  the new and old bases are connected as
follows
\[  z_l=\sum_{s=1}^n c^s_l y_s \ \ \ (l=1, \ldots, n) \]

\section{Metric in synthetic differential geometry}
In this section we shall define metric concepts
within a context of SDG.
\vspace{3\baselineskip}

\subsection{The metric properties of $R^n$}
In this paragraph we shall define metric concepts on $R^n$.
\begin{dfn}
{\rm
 A map $(\cdot ,\cdot):R^n\times R^n \rightarrow R$
 which satisfies the following conditions:
 \begin{enumerate}
   \item $\bar{x} \# 0 \Rightarrow \exists \bar{y} \in R^n: $
              $ (\bar{x}, \bar{y} )\# 0 $\\
	 $\bar{x} =0   \Rightarrow  (\bar{x} , \bar{x} )=0 $
  \item $(\bar{x}, \bar{y})=(\bar{y}, \bar{x}) $
  \item $(\bar{x} +\bar{y} ,\bar{z})=(\bar{x} ,\bar{z}) +(\bar{y} ,\bar{z}) $
  \item $(\lambda \cdot \bar{x} ,\bar{y})=\lambda \cdot (\bar{x} ,\bar{y}) $
 \end{enumerate}
 where $\lambda \in R$\ , \ $\bar{x}, \bar{y}, \bar{z}\in R^n$,
 will be called a {\it scalar product} on $R^n$.
}
\end{dfn}
So as $R$ is a Pythagorean (\ref{equ_Pyth})   and a formal real
 (\ref{equ_freal}) ring
we may define a norm of  vector as  follows
\begin{dfn}
 {\rm
   Let $\bar{x}\in R^n$ such that $\bar{x}\# 0 $.Then a number
   $ \| \bar{x} \|=\sqrt{(\bar{x} ,\bar{x})}$.
   will be called a {\it norm} of the vector $\bar{x}$.
}
\end{dfn}
\begin{dfn}
{\rm
  We shall speak, that the vectors
  $\bar{x} ,\bar{y} \in R^n$, such that $\bar{x} \# 0, \bar{y} \# 0$,
  are {\it orthogonal} \ if $(\bar{x}, \bar{y})=0$.
}
\end{dfn}

We shall call the $R^n$ with a scalar product $(\cdot, \cdot)$
as  Euclidean space if
$(\bar{x}, \bar{x} ) > 0$ for all $\bar{x} \# 0$ and
as pseudo-Euclidean if $(\bar{x}, \bar{x} )$ may be
both positive and negative.

Let  $\{\bar{e}_1, \ldots, \bar{e}_n\}$ be a basis of $R^n$ .
We shall denote
$g_{ij}=(\bar{e}_i, \bar{e}_j) \ \ \ (i,j=1, \ldots, n)$,
The scalar product $\bar{x}, \bar{y}\in R^n$ can
be recorded as
\[ (\bar{x} ,\bar{y})=\sum_{i,j=1}^n g_{ij}\cdot x^i y^j. \]
\begin{teo}\label{teo_det}
  Determinant of the matrix $\{ g_{ij}\}$, apart from a zero.
\end{teo}
\begin{prv}
 We shall determine the following linear forms by formulas:
 \[ \ell^i (\bar{x})=(\bar{e}_i ,\bar{x}). \]
 Let us show, that $\ell^i (\bar{x}) $ are mutually free.
 For this purpose is enough to show that the form $\ell(\bar{x})$,
 defined as the linear  combination
 \begin{equation}\label{equ_sum}
   \ell (\bar{x})=\sum_{i=1}^n \lambda_i \ell^i (\bar{x})
 \end{equation}
 where at least one $\lambda_i \# 0$,  is aparted from a zero.
 By linearity we can write
 \[ \ell(\bar{x})=(\bar{a} ,\bar{x}) \]
 where $\bar{a}=\sum_{i=1}^n \lambda_i \bar{e}_i.$
 In a force of the definition of $\lambda_i $ we have that $\bar{a}$
 apart from a zero,
 and consequently $(\bar{a} ,\bar{a}) \# 0 $.
 Thus
 \[ \ell (\bar{a}) \# 0 \]
 From (\ref{equ_fild})  may be deduce that exists $\bar{e}_i$ such, that
 \[ \ell (\bar{e}_i) \# 0.\]
 Hence $\ell (x) $ is aparted from a zero.
 It means, that $\ell^i (\bar{x}) $ are mutually free and, under the
 Theorem \ref{teo_Rank} \[ \det{ \{ g_{ij}\} }\# 0.\] $\Box $
\end{prv}

\subsection{Tangent bundle of $R^n$}
As in \cite{Kock_SDG} we shall assume that a
tangent bundle to $R^n$ is  the object ${R^n}^D$ (exponential object).
Let us denote it as $TR$.
>From Axiom 1  follows that $TR \cong R^n\times R^n$,
whence a tangent vector to $R^n$ in a point $a=(a_1, \ldots, a_n) $
is a map $t:D\rightarrow R^n$ of the form
\[ t(d)=(a_1+d\cdot b_1, \ldots, a_n+d\cdot b_n) \]
where $\bar{b}=(b_1, \ldots, b_n) \in R^n$.
The {\it main part} $\gamma:TR\rightarrow R^n $ is a map such that
$\gamma (t)=\bar{b}$.It establishes isomorphism of $R$-modules
$T_aR^n \stackrel{\gamma}{\cong}R^n$.
\begin{dfn}
{\rm
  We shall speak, that a vector $t\in TR^n$ is {\it apart from a
  zero} ($t\#0 $), if the main part $\gamma (t) $ apart from a zero in $R^n$.
}
\end{dfn}
We shall give definition of scalar product in $T_aR^n$.
\begin{dfn}
{\rm
  As {\it  scalar product} of two tangent  vectors to $R^n$ in a point
  $a$ we shall name scalar product of their main parts in $R^n$, that is
  \[ <\cdot ,\cdot>:TR^n \times_{R^n}TR^n
     \stackrel{\gamma \times \gamma}{\longrightarrow}R^n \times R^n
     \stackrel{(\cdot, \cdot)}{\longrightarrow}.
  \]
  Let $\|t\|=\sqrt{< t, t >}$ be a {\it norm(module)}
  of a vector $t$, such, that $t\#0 $.
}
\end{dfn}
\begin{dfn}
{\rm
   We shall name a map $c:[a,b] \rightarrow R^n$
   as {\it curve}  on  $R^n$.
}
\end{dfn}
\begin{dfn}
{\rm
 The tangent vector $\dot{c}(t):D\rightarrow R^n$, such that
  \[ \dot{c} (t)(d)=c(t+d)=(c_1 (t+d,\ldots, c_n (t+d) \]
  \[ =(c_1 (t), \ldots, c_n (t))+d\cdot (c_1'(t),\ldots ,c_n'(t))=
    c(t)+d\cdot c' (t).\]
   we shall name as
  {\it speed vector} of a curve $c$ in a point $t\in [a,b]$}
\end{dfn}

The module of a speed vector
defines a map $\| \dot{c}\|:[a,b] \rightarrow R $
\[ \| \dot{c} (t) \|=\sqrt{< \dot{c}(t), \dot{c}(t) >} \]
\begin{dfn}
{\rm
  A {\it length of a curve} $c:[a,b]\rightarrow R^n $
  such, that $\dot{c}(t) \# 0\ \ \ \forall t\in [a,b] $,
  given on a interval $[a, b]$ such, that $a\leq b$ is
  integral from a module of its  speed vector, i.e.
  $\ell (c)=\int\limits_a^b \| \dot{c}(t) \| dt $.
}
\end{dfn}
\begin{exm}
 Let $f:[a,b] \rightarrow R$ where  $a\leq b$. We shall find
 a length of the curve  $c(t)=(t,f(t))$ which is a graph of the function $f$.
 We have $\dot{c}(t)=(t, f(t))+(d, d\cdot f'(t)) $, hence
 length $\ell(c)=\int\limits_a^b \sqrt{1+f'^2 (t)}dt$.\\
 Notice, that in this case
 $\dot{c}(t) \# 0\ \ \ \forall t\in [a,b] $
	because $1^2+f'^2 (t) \# 0$.
\end{exm}
\begin{prp}
{\rm The length of a curve does not depend from parametrization.}
\end{prp}
\begin{prv}
 Follows in a standard manner from properties of  replacement of variables in
 integral.$\Box $
\end{prv}

\subsection{An element of  curve's arch}
In this paragraph we shall deduce the classical formula for differential
of a element of  curve's arch  on the plane $R\times R$.

At the beginning we shall give the following definition:
\begin{dfn}
{\rm
  Let $M$ be an arbitrary  object in \E\, and $f:M\rightarrow R$.
  The {differential} of $f$ is the  composition
  \[ df:M^D\stackrel{f^D}{\longrightarrow}R^D
           \stackrel{\gamma}{\longrightarrow}R,  \]
  where $\gamma$ is the main part.
}
\end{dfn}

Let us consider a curve $c:[0,1] \rightarrow R $.
In coordinates it is $c(t)=(x(t), y(t)) $.
We shall assume, that a speed vector
$\dot{c}(t) \# 0\ \ \ \forall t\in [0,1] $.
We have
 \[ \dot{c}(t)=c(t+d)=c(t) + d\cdot c'(t),\]
where $c'(t)=(x'(t), y'(t))$.
The length of the curve $c$ will be  equal to
\[ \ell(c)=\int\limits_0^1 \sqrt{\| \dot{c}(\tau) \| }
 d\tau=\int\limits_0^1 \sqrt{x'^2 (\tau)+y'^2 (\tau)}d\tau. \]
We shall define a {\it length of  curve's arch} $s:[0,1] \rightarrow R$ as
\[ s(t)=\int\limits_0^t \sqrt{x'^2 (\tau)+y'^2 (\tau)}d\tau. \]
On property of integral with a variable bound we have
\[ s'(t)=\sqrt{x'^2 (t)+y'^2 (t)}. \]
Hence, we receive, that
\[ s(t+d)=s(t)+d\cdot \sqrt{x'^2 (t)+y'^2 (t)}\ \ \ \forall d\in D \]

Now let  us consider differential of  $s(t)$. On definition it
is a map
\[ds:[0,1]^D\stackrel{s^D}{\longrightarrow}R^D
   \stackrel{\gamma}{\longrightarrow}R.
\]
\[ ds(a+d\cdot b)=\gamma(s(a+d\cdot b)=\]
\[=\gamma (s (a)+d\cdot b\sqrt{x'^2 (t)+y'^2(t)})
  =b\cdot \sqrt{x'^2 (t)+y'^2 (t)} \]

Similarly for differentials $dx$ and $dy$ we have:
\[ ds(a+d\cdot b)=b\cdot \sqrt{x'^2 (t)+y'^2 (t)} \]
\[ dx(a+d\cdot b)=b\cdot x' (t)  \]
\[ dy(a+d\cdot b)=b\cdot y' (t)  \]
Hence, using operations of addition and multiplication of
functions from $R^{[ 0,1 ]}$, we receive the classical formula for
differential of a element of  curve's arch:
\begin{equation}
  ds^2=dx^2+dy^2
\end{equation}
Having conducted replacement of coordinates $x(u,v) ,y(u,v)$, by similar
reasons, it is possible to show, that
\[ dx=\frac{\partial x}{\partial u}\cdot du+
   \frac{\partial x}{\partial v}\cdot dv \]
\[
 dy=\frac{\partial y}{\partial u}\cdot du+
 \frac{\partial y}{\partial v}\cdot dv  \]
and to deduce the formula:
\begin{equation}
 ds^2=E\cdot du^2+2 F\cdot du dv+G\cdot dv^2
\end{equation}
\[ E=(\frac{\partial x}{\partial u})^2 +(\frac{\partial y}{\partial u})^2 \]
\[ F=\frac{\partial x}{\partial u}\cdot\frac{\partial x}{\partial v}+
     \frac{\partial y}{\partial u}\cdot\frac{\partial y}{\partial v}  \]
\[ G=(\frac{\partial x}{\partial v})^2+(\frac{\partial y}{\partial v})^2  \]

\subsection{Riemannian structure on formal manifold}
A notion of formal manifold in SDG is a generalization of a classical
notion of a $C^\infty$-manifold.
In this paragraph we shall show, that on formal manifold it is possible to
develop a Riemannian geometry.

Let $M$ be a $n$-dimensional formal manifold\cite{Kock_SDG}
and $\{U_i\stackrel{\varphi_i}{\longrightarrow}M\}$ be a cover of $M$ by
formally etale monomorphisms, where $U_i$ are  model objects,
i.e. formal etale subobjects in $R^n$.
The pair $(U_i,\varphi_i) $ will be called {\it a local card} on $M$.

So as $\varphi $ is monomorphism, it is convertible on the image
$\varphi (U) $, and,
hence, $\varphi^{-1}:\varphi (U) \rightarrow U $  is determined.

We shall consider tangent bundle $TM=M^D$.
As $M$ is a formal manifold, is valid, that  $T_{p}M\cong R^n$
for each $p\in M$.

Let $v:D\rightarrow M$ be a tangent vector to $M$ in a point $p$ and
$(U,\varphi) $ be a local card such, that $\varphi^{-1}(p)=0$.
In this case $\varphi^{-1}\circ v$ is a tangent vector to $U$ in $0$.
Since  $U$ is subobject of $R^n$ it is valid that $TU\cong U\times R^n$
and, hence, the vector can be recorded as
$\varphi^{-1} \circ v(d)=(0,\ldots ,0)+d\cdot (v_1, \ldots, v_n)$,
where $(v_1,\ldots ,v_n)\in R^n$.

We shall denote through $\partial_i$ vectors
\[ \partial_i(d)=(0, \ldots , 0)+d\cdot (0, \ldots ,
                                         \stackrel{i}{1}, \ldots , 0) \]
The vectors $\partial_i \circ \varphi$ will form a basis of $T_{p}M$.
\begin{dfn}
{\rm We shall speak, that vectors $u, v\in T_{p}M$
  are {\it aparted } if $(\varphi^{-1}\circ u) \# (\varphi^{-1}\circ v) $
  in $T_{0}U$.
}
\end{dfn}
\begin{dfn}\label{dfn_Riem}
{\rm
  Let $M$ be a formal manifold.
  A map $g:TM\times_M TM\rightarrow R $
  will be called {\it a metric tensor (a Riemannian structure)}
  on $M$ if following conditions are executed.
  \begin{enumerate}
   \item  $v\# 0 \Rightarrow \exists u: g(v, u)\# 0 $\\
	  $v=0   \Rightarrow g(v, v)=0 $
    \item $g(v, w)=g(w, v) $
    \item $g(u+v, w)=g(u, w)+g(v, w) $
    \item $g(\lambda \cdot v, w)=\lambda \cdot g (v, w) $
  \end{enumerate}
  where $\lambda \in R$, $v, w, u \in TM $ so that $v(0)=w(0)=u(0)$.
}
\end{dfn}
\begin{dfn}
{\rm
  Let $v\in TM $ such, that $v\# 0$. Then  a number $\| v\| =\sqrt{g(v,v)}$
  will be called a {\it  norm} of a vector $v$.
}
\end{dfn}

We shall call the M with a metric tensor $g$
as  Riemannian space if
$g(v, v) > 0$ for all $v \# 0$ and
as pseudo-Riemannian if $g(v, v)$ may be
both positive and negative.

Let us consider a tangent space $T_{p}M$ for some $p\in M$.
Then a map $g^p:T_{p}M\times T_{p}M\rightarrow R $ defined
as $g^{p}(u,v)=g(u,v)$, for $u, v\in T_{p}M$, is
a scalar product on $T_{p}M$.

We shall denote \[ g^p_{ij}=g^p(\partial_i ,\partial_j).\]
In a force of the Theorem \ref{teo_det} we have that $det{(g^p_{ij})}\# 0 $.

For any $u, v\in T_{p}M $ we have
\[ g^p (u, v)=g^p_{ij}\cdot u^i v^j, \]
where $ u^i, v^j$ are coordinates of vectors $u, v$ at decomposition on basis
$\{ \partial_i \}$.

As well as in case of $R^n$ it is possible to define a curve on $M$
as a map $c:[a,b] \rightarrow M$, with a speed vector in
a point $t\in [a,b] $ equal to $\dot{c}(t)(d)=c(t+d)$.

We shall assume, that $a\leq b$ and
$\dot{c}(t) \# 0 \ \ \ \forall t\in [a,b]$.Then it is possible to
define a length $\ell(c)$ of a curve $c(t)$ as
\[ \ell (c)=\int\limits_a^b \sqrt{g (\dot{c}(t), \dot{c}(t))}dt \]
\begin{note}
 It is interesting to note, that in general we can't define
 on $M$ internal metric $\rho $ as
 \[ \rho (p, q)=\inf_{c p\frown q}{\ell (c)}. \]
  The reason is that $R$ is not
  order complete, and therefore the existence  of
  $\inf_{c p\frown q}{\ell (c)}$ needs to be proved positive.
\end{note}

\subsection{Models of Riemannian  structures on formal manifolds}
The well adapted model\cite{Kock_SDG} of SDG is in such category \E\
that exist functor $i:Mf\rightarrow$ \E\  from a category of
$C^\infty$-manifolds to \E, which allow to compare a classical
differential geometry with a synthetic  one.
In this paragraph we shall show, that in well adapted models
exist a Riemannian structure on formal manifolds of the kind $M=i({\cal M})$.

Let us consider a well adapted model
$i:Mf\rightarrow$\E, there $Mf$ is a category of $C^\infty$-manifolds.
Let $\cal R$ be the field of real numbers with a natural
Riemannian  structure on it
and $\cal M$ be a $C^\infty$-manifold  with given on it  a
Riemannian structure $g:T{\cal M}\times_{\cal M}T{\cal M}\rightarrow{\cal R}$
\begin{enumerate}
  \item $v\# 0 \Rightarrow \exists u: g(v, u)\# 0 $\\
        $v=0   \iff        g(v, v)=0 $
 \item $g(v,w)=g(w,v) $
 \item $g(u+v, w)=g(u,w)+g(v,w) $
 \item $g(\lambda \cdot v, w)=\lambda \cdot g(v,w) $
\end{enumerate}
Where $\lambda \in \cal R$, and $v, w, u \in T{\cal M}$
	such that $v(0)=w(0)=u(0) $.

We shall assume, that $g$ is $C^\infty$-mapping. It means, that
\[ g\in Hom_{Mf}(T{\cal M}\times_{\cal M}T{\cal M},{\cal R}). \]

We shall consider
$i(g):i(T{\cal M} \times_{\cal M} T{\cal M}) \rightarrow i({\cal R}) $.
The diagram
\begin{eqnarray*}
  T\cal M\times_{\cal M} T\cal M & \longrightarrow & T\cal M         \\
  \downarrow                     &                 & \downarrow \pi  \\
  T\cal M                & \stackrel{\pi }{\longrightarrow} & \cal M
\end{eqnarray*}
 is transversal pull back in $Set$, hence, in a force of
Axiom A  of well adapted models\cite{Kock_SDG},
it is preserved by $i$, i.e
\[  i(T{\cal M}\times_{\cal M}T{\cal M})=i(T{\cal M})
 \times_{i({\cal M})} i(T{\cal M}). \]

We shall denote $M=i({\cal M}), R=i({\cal R}) $.
It is known that $M$ is a formal manifold.
By virtue of Axiom C  of well
adapted models\cite{Kock_SDG} we have that
$i(T{\cal M}) \cong Ti({\cal M})=TM$.
	  In a result we receive, that
\[ i(g):TM\times_M TM\rightarrow R \]

The conditions (1) - (4) of the Riemannian structure $g$ on $\cal M$ can be
expressed in the form of commutativity of the appropriate diagrams in $Mf$.
Functor $i:Mf\rightarrow$ \E\  save them and, hence, the
conditions of Definition \ref{dfn_Riem} of a Riemannian structure on formal
manifold will be executed for $i(g)$.

Thus we have shown, that on formal manifolds of kind $i({\cal M})$, in
well adapted models, exist  a Riemannian structure of kind $i(g)$.

\section{Einstein's equations of a field in SDG}
 In this section we show that it is possible to write
 Einstein's equations of a field in SDG. For this we need in some
new notions of SDG.

\subsection{Connection in SDG}
In this paragraph we shall give a notion and some properties
of connection in SDG. This results are taken from \cite{KockReyes_Conn}.

First of all let us make note about tensors  in SDG.
For any  $R$-modules $U$ and $V$ we can use a classical definition of
qtensor product \cite{Kasch}. For this definition
 all algebraic properties will be valid and all  algebraic
operations will be definable.

Hence they will be valid for $R^n$.
Moreover, from the Theorem \ref{teo_det}
follows that the operations of rising and lowering indexes are
definable  for $R^n$ too.

Let us see an object $M$  in \E.
We define an infinitesimal object $D \lor D$ as
\[ D \lor D=\{(x, y)\in R\times R\ |\  x\cdot y=0 \} \]
 It easy to see that
\[ D\lor D \subseteq D\times D \subseteq R\times R.\]
We denote the inclusion $D\lor D \subseteq D\times D$ as $j$.

Let us see the object $M^{D \lor D}$. For infinitesimal
linear object \cite{Kock_SDG} $M$ ( for a example for formal manifold )
we have
\[ M^{D \lor D} \cong M^D \times_M M^D.\]
So we can see the elements of $M^{D \lor D}$ as a "crosses" of
tangent vectors in each point of $M$.

Let ${M^j}$ be the restriction map
\[ M^{D\times D} \stackrel{M^j}{\longrightarrow} M^{D\lor D}.\]
Then we give the following
\begin{dfn}
 {\rm
     The connection $\nabla$ on a tangent bundle
     $M^D \stackrel{\pi}{\longrightarrow} M$ of the object $M$
     is a section of the restriction map $ M^j$, i.e
   $ M^{D\lor D} \stackrel{\nabla}{\longrightarrow} M^{D\times D}$.
 }
\end{dfn}
Geometrically, this definition may be understood as
a complementation of a "cross" to a infinitesimal "net" (an element
of the object $M^{D\times D} $ ) or as parallel transport of
the second vector along the infinitesimal segment of the line
given by the first vector ( elements of the $ M^{D\lor D} $).

Bases on this definition in \cite{KockReyes_Conn} are
given a definition of curvature $k$ of a connection $\nabla$
on a tangent bundle $\pi:M^D\rightarrow M$ as a map
\[ k:(M^D)^{D\times D}\longrightarrow M^D .\]

In a case than $M$ is a etale subobject $U\rightarrow R^n$  in $R^n$
it is possible to write a connection  on tangent bundle
$\pi:U^D\rightarrow U $ in coordinates\cite{KockReyes_Conn}.
More exact, connection $\nabla$ became a map
\[ U^D\times_U U^D \cong U\times R^n\times R^n
\stackrel{\nabla}{\longrightarrow}
   U\times R^n\times R^n\times R^n \cong U^{D\times D}.\]
Sine connection is a section of the map $M^j$, which in coordinates has form
\[ (u,v_1,v_2,v_3)\longmapsto (u,v_1,v_2),\]
it can be written as
\[ \nabla(u,v_1,v_2) = (u,v_1,v_2,\bar{\nabla} (u,v_1,v_2)).\]

 If $\nabla$ is affine connection\cite{KockReyes_Conn} then
 $\bar{\nabla} (u,v_1,v_2))$ is bilinear  from  $v_1,v_2$. And
hence, can be defined by 3$n$ indexed family of functions
$\Gamma_{ij}^k:U\rightarrow R$.

If $u\in U$ and $\{e_i\}$ is the canonical base in $R^n$
then, as it show in \cite{KockReyes_Conn}, the curvature $k$
of a connection  $\nabla$ has form
\[
 R_{kij}^l = \frac{\partial}{\partial x^i} \Gamma_{jk}^l (u) -
             \frac{\partial}{\partial x^j} \Gamma_{ik}^l (u) +
 \sum_{\alpha} \Gamma_{ik}^{\alpha} (u)\cdot \Gamma_{j \alpha}^l (u) -
	       \Gamma_{jk}^{\alpha} (u)\cdot \Gamma_{i \alpha}^l (u),
\]
where $\Gamma_{ij}^k (u)$ is a $k$-th coordinate of
 $\bar{\nabla} (u,e_i,e_j)$.
This formula is equivalent to the classical one.

\subsection{Riemann -- Christoffel's tensors}
In this paragraph we define a tensor of curvature on a formal
manifold with a metrical tensor.
Let us see a formal manifold $M$ with a metrical tensor $g$.
Since each  local card  of $M$ is a etale subobject  of $R^n$
we can define connection on it by given $\Gamma_{ij}^k$.

Let $U$ be a local card. We define $\Gamma_{ij}^k$ (Christoffel's
symbols of a second kind) in a classical manner using a coefficients
of metrical tensor.
 For this we define Christoffel's
symbols of a first kind by formulas
\[ [ij, k] \equiv \frac{1}{2} \left( \frac{\partial g_{ik}}{\partial x^j} +
   \frac{\partial g_{jk}}{\partial x^i} - \frac{\partial g_{ij}}{\partial x^k}
     \right)\ \ \ (i, j, k = 1,\ldots ,n), \]
where $g_{ij}$ are coefficients of metrical tensor in $U$.

And then  we define $\Gamma_{ij}^k$ by formulas
\[ \Gamma_{ij}^k  \equiv g^{k\alpha} [ij, \alpha],\]
where  $g^{k\alpha}$
are coefficients of contvariant metrical tensor in $U$.

So we  define the connection all local cards. Hence we
define connection $\nabla$ on $M$.

The coefficients  $R_{kij}^l$ of the curvature  $k$ defines
co called Riemann -- Christoffel's tensors of the second kind.
Lowering down indexes we shall receive associated tensor
\[ R_{ijkl} \equiv g_{i\alpha} R_{jkl}^\alpha, \]
Riemann -- Christoffel's tensors of the first kind.

>From the definition of these tensors follows that
they possessed of all classical properties of symmetry under the
indexes change.

\subsection{Einstein's equations}
Let us see 4 dimension pseudo-Riemann formal  manifold $M$ with
metric tensor $g$ and let $\nabla$ be the connection on $M$
constructed as above.

Using the tensor operations we may define Ricci's  tensor $R_{ij}$ by
formula  $R_{ij} = R_{ij\alpha}^\alpha$  and Einstein's
tensor by formula $ G_j^i \equiv R_j^i - \frac{1}{2} \delta_j^i R$,
where $R \equiv g^{ij} R_{ij}$ and $R_j^i = g^{ki} R_{kj}$.

Having the definitions of this tensors me may write the
Einstein's equations of a field as
 \[ - \kappa T_{ij} = R_{ij} - \frac{1}{2} R g_{ij},\]
where $T_{ij}$ is tensor of energy-impulse and $\kappa$ is a constant.

So we have shown that SDG may be viewed as base for
consideration of general theory of relativity. Particular it gives
an ability to construct an intuitionistic models of
general relativity in a toposes which are the well adapted models for SDG.


\begin{thebibliography}{99}
 \bibitem{Kock_SDG}{\bf A.Kock}, Synthetic Differential Geometry,
  Cambridge University Press, 1981.
 \bibitem{Kock_FRR}
  {\bf A.Kock}, Formal real rings, and infinitesimal stability, in Topos
  Theoretic Methods in Geometry, Aarhus Math. Inst. Var. Publ. Series No 30
  (1979)
 \bibitem{KockReyes_Conn} {\bf A.Kock, G.E.Reyes}, Connections in formal
  differential geometry,  in Topos Theoretic Methods in Geometry,
  Aarhus Math. Inst. Var. Publ. Series No 30 (1979)
 \bibitem{Lawvere_CD}{\bf F.W.Lawvere}, Categorical Dynamics, in
  Topos Theoretic Methods in Geometry, Aarhus Math. Inst. Var. Publ.
  Series No  30 (1979)
 \bibitem{Lawvere_ETCS}{\bf F.W.Lawvere}, An elementary
  theory of the category of sets, Proc. Nat. Acad. Sci. USA 52 (1964), 
  1506 -  1511.
 \bibitem{McLarty}{\bf C.McLarty}, Local, and some global results in
  synthetic differential geometry, in Topos Theoretic Methods in Geometry,
  Aarhus Math. Inst. Var. Publ. Series No 35 (1983)
 \bibitem{Mulvey_IA}{\bf C.Mulvey}, Intuitionistic algebra and
   representation of rings, Memoris AMS 148 (1974).
 \bibitem{Heyting_I}{\bf A.Heyting}, Intuitionism, Amsterdam, 1956.
 \bibitem{Lang_A}{\bf S.Lang}, Algebra, New York, 1965.
 \bibitem{Kasch}{\bf F.Kasch}, Moduln und ringe, Stuttgart, 1977.
 \bibitem{Kobayashi_DG}{\bf S.Kobayashi, K.Nomizu}, Foundations
  of differential geometry, New York, London, Intercience publishers,1963.
\end{thebibliography}
\end{document}